# Journal Name

## ARTICLE TYPE

# Structural correlations in highly asymmetric binary charged colloidal mixtures

Elshad Allahyarov,[a,b,c] Hartmut Löwen,[b] and Alan Denton[d]

We explore structural correlations of strongly asymmetric mixtures of binary charged colloids within the primitive model of electrolytes considering large charge and size ratios of 10 and higher. Using computer simulations with explicit microions, we obtain the partial pair correlation functions between the like-charged colloidal macroions. Interestingly the big-small correlation peak amplitude is smaller than that of the big-big and small-small macroion correlation peaks, which is unfamiliar for additive repulsive interactions. Extracting optimal effective microion-averaged pair interactions between the macroions, we find that on top of non-additive Yukawa-like repulsions an additional shifted Gaussian attractive potential between the small macroions is needed to accurately reproduce their correct pair correlations. For small Coulomb couplings, the behavior is reproduced in a coarse-grained theory with microion-averaged effective interactions between the macroions. However, the accuracy of the theory deteriorates with increasing Coulomb coupling. We emphasize the relevance of entropic interactions exerted by the microions on the macroions. Our results are experimentally verifiable in binary mixtures of micron-sized colloids and like-charge nanoparticles.

## 1 Introduction

Charged colloidal suspensions are interesting model systems for classical many-body systems as their effective interactions can be tuned and tailored externally by adding depletants[1], salt and changing the solvent polarity[2–8]. From the early days of colloidal fluids, light scattering data of charged suspensions[9,10] have been used to test liquid integral equations theories[11–14] which predict the pair correlation between the colloidal particles based on their effective screened Coulomb (or Yukawa) pair potential[15], or to compare to computer simulations using the microion-resolved primitive model of electrolytes[16–20].

This can subsequently generalized to binary mixtures of colloidal suspensions with different charges[21–37]. In this case, the effective interaction forces and particle correlations can be determined experimentally as well[22–24,38] and compared to predictions of model simulations[39–45] or theories[46–64]. In general, a purely repulsive mixture is non-additive characterized by a non-additivity parameter $\Delta$[24,65], the sign of which implies whether there is clustering of similar species or microphase separation.

In most of the theoretical and simulational studies so far, binary mixtures of charged particles were assumed to differ not too much in charge and size[65,66]. This is realized for many[23,24] but not for all charged colloidal mixtures (see[67] for an experimental example). Even more importantly, when micron-sized colloidal particles are mixed with nano-particles[34,36], a strongly asymmetric mixture arises. For highly asymmetric mixtures, correlations are much harder to predict by theory as liquid integral equations typically break down for large asymmetries. Also in simulations, the length scale separation between big and small macro-particles reduces sampling drastically such that a simulation much more involved in the asymmetric case. When taking explicit microions into account within the primitive model approach of electrolytes, the situation is even more challenging since there are two length scale gaps involved, the first between the colloids and the nanoparticles and the second between the nanoparticles and the microions.

In this paper we present a comprehensive computer simulation study using the highly asymmetric primitive model of electrolytes for strongly asymmetric colloidal suspensions. By computing the partial pair distribution functions in the fluid binary mixture, we find that the cross-interaction big-small correlation peak amplitude is smaller than that of the big-big and that of the small-small correlations. This is uncommon for almost symmetric repulsive mixtures. Moreover we show that entropic forces arising from the excluded volume interactions due to the finite core size contribute significantly to the total interaction forces. We then extract optimal effective microion-averaged pair potentials between the macroions following a scheme proposed for one-component systems[68]. As a result, we find that non-additive Yukawa-like repulsions provide a good fit but an additional shifted attractive Gaussian potential between the small macroions is needed to reproduce the correct pair correlations. It is suggested that this additional potential originates from the effective attraction between the small macroions located in the cage created by the macroions, mediated by the screening counterions between the neighboring macroions. For small Coulomb couplings, the behavior is reproduced in a coarse-grained theory which was proposed by one of us[69–71]. However, the theory deteriorates for larger Coulomb coupling. Our results are experimentally verifiable in binary mixtures of micron-sized colloids and like-charge nanoparticles.

[a] *Theoretical Department, Joint Institute for High Temperatures, Russian Academy of Sciences (IVTAN), 13/19 Izhorskaya street, Moscow 125412, Russia; E-mail: elshad.allahyarov@case.edu*
[b] *Institut für Theoretische Physik II: Weiche Materie, Heinrich-Heine Universität Düsseldorf, Universitätstrasse 1, 40225 Düsseldorf, Germany*
[c] *Department of Physics, Case Western Reserve University, Cleveland, Ohio 44106-7202, United States*
[d] *Department of Physics, North Dakota State University, Fargo, ND 58108-6050 United States*

The paper is organized as follows. In section 2 we describe the details of our primitive model simulations for the binary colloidal system. The results obtained for the partial pair correlation functions are discussed in section 3 and compared to the prediction of the coarse-grained theory. In section 4 we explore the role of entropic forces in the macroion interactions. Section 5 is devoted to the extraction of the optimal pairwise interactions between the macroions. We conclude in section 6.

## 2 Details of the Primitive Model

We consider a three-component binary colloidal suspension consisting of $N_Z$ big macroions of charge $q^{(Z)} = Ze$ and size $\sigma_Z = \sigma$ at positions $\vec{r}_i^{(Z)}$ ($i=1,...,N_Z$), $N_z$ small macroions of charge $q^{(z)} = ze$ and size $\sigma_z = \sigma/10$ at positions $\vec{r}_j^{(z)}$ ($i=1,...,N_z$), and $N_c = ZN_Z + zN_z$ monovalent counterions of charge $q^{(c)} = -e$ and size $\sigma_c = \sigma/600$ at positions $\vec{r}_\ell^{(c)}$ ($\ell=1,...,N_c$). Here $e$ is the absolute value of the electron charge. We fix the size ratio to reduce parameter space to realistic values. The pair interaction potential between the species $\alpha$ and $\beta$ with $\alpha, \beta \in \{Z,z,c\}$ are given as a combination of excluded volume and Coulomb interactions (in SI units),

$$V^{(\alpha\beta)}(r_{ij}) = \begin{cases} \infty, & \text{for } r_{ij} \leq \sigma_{\alpha\beta} \\ q^{(\alpha)}q^{(\beta)}/(4\pi\varepsilon_0\varepsilon r_{ij}), & \text{for } r_{ij} > \sigma_{\alpha\beta} \end{cases} \quad (1)$$

where $\vec{r}_{ij} = \vec{r}_i^{(\alpha)} - \vec{r}_i^{(\beta)}$ with $i \in 1,...,N_\alpha$ ($\alpha = Z,z,c$) and $j \in 1,...,N_\beta$ ($\beta = Z,z,c$) is the distance between the two particles, $\sigma_{\alpha\beta} = (\sigma_\alpha + \sigma_\beta)/2$ is their additive hard core diameter, $\varepsilon_0$ is the vacuum permittivity, and $\varepsilon$ is the relative permittivity of the suspension. For simplicity, we assume that $\varepsilon$ is the same throughout the system in order to avoid image charge and dielectric boundary effects.

The following parameters characterize the intensity of interparticle interactions and counterion screening effects in binary colloidal systems:
- the packing fraction of big macroions, $\eta$,
- the packing fraction of small macroions, $\eta_s$,
- the Debye-Hückel inverse screening length

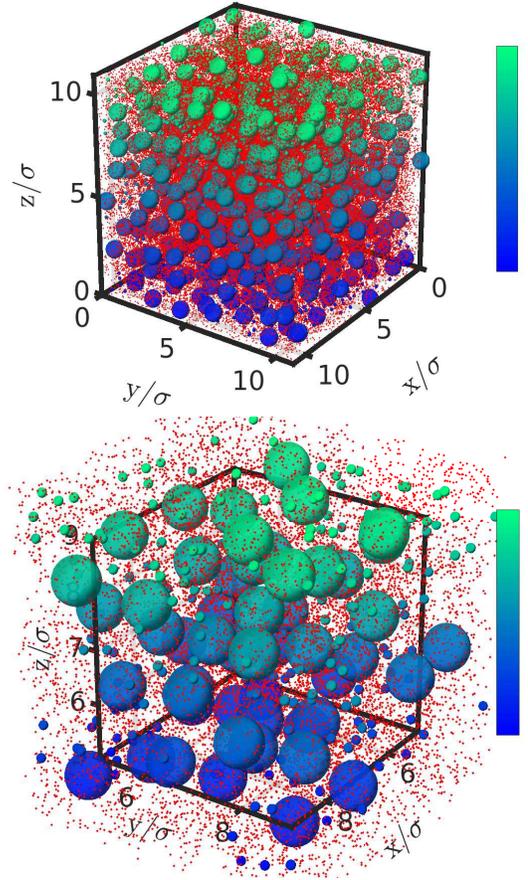

Fig. 1 (Color in online) Simulation snapshots from the run B2 in Table 1. Left picture- a full view, right picture- a zoomed view. Big and small macroions are shown as spheres of relevant diameters, counterions are shown as scattered red dots. A color gradient from blue to green on the colorbar corresponds to the macroion altitude in the cell.

$\kappa = \sqrt{n_c e^2/(\varepsilon_0 \varepsilon k_B T)}$ of counterions, where $k_B$ is the Boltzmann constant and $T$ is the temperature in the system,
- the Bjerrum length, $\lambda_B = e^2/(4\pi\varepsilon_0\varepsilon k_B T)$,
- the average distance between the big macroions, $a = L\left(6/(\pi N_Z)\right)^{1/3}$, where $L$ is the edge length of the cubic simulation box,
- the average distance between the small macroions, $b = L\left(6/(\pi N_z)\right)^{1/3}$,
- the Coulomb coupling parameter between the big macroions, $\Gamma = Z^2 \exp(-\kappa a)\lambda_B/a$,
- the Coulomb coupling parameter between the small macroions, $\xi = z^2 \exp(-\kappa b)\lambda_B/b$.

Note that, the cell volume accessible to counterions is $V = L^3(1 - \eta - \eta_s)$, thus the available-volume corrected counterion density and the inverse screening length become $n_c/(1 - \eta - \eta_s)$ and $\kappa/\sqrt{(1 - \eta - \eta_s)}$, respectively. Additionally, for concentrated colloidal systems considered in this work, the true electrostatic screening length might strongly differ from the classical Debye-Hückel length $1/\kappa$, see for details Ref.[72].

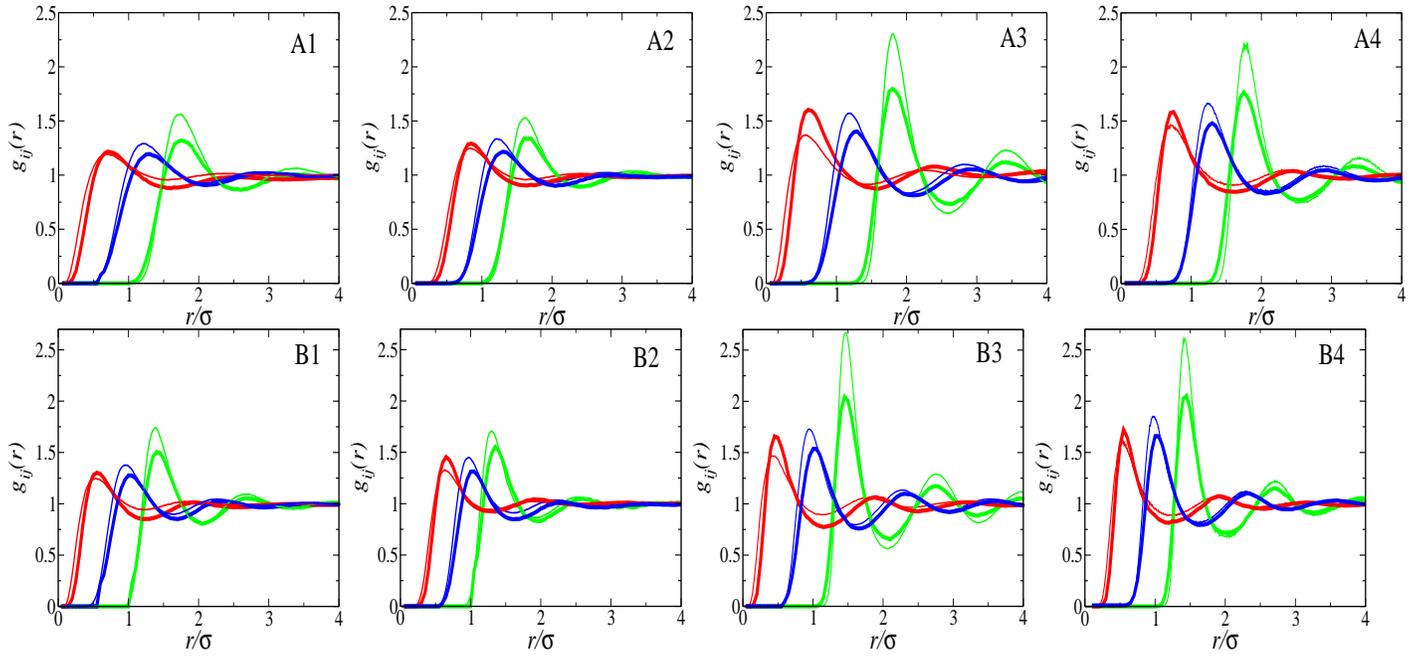

Fig. 2 (Color in online) Pair correlations $g_{ij}(r)$ for the runs A1-A4 (first row) and for the runs B1-B4 (second row). Thin lines are for the CGM, thick lines are for the PM simulations. Green lines are for $i=j=Z$, red lines are for $i=j=z$, and blue lines are for $i=Z$ and $j=z$.

Table 1 Primitive model simulation parameters for different runs. The quantities listed in the first row are explained in the text. The packing fraction $\eta_s$ for small macroions was 0.01 for run D1, and 0.001 for run D2.

| Runs | $Z$ | $z$ | $\eta$ | $N_c$ | $n_c\sigma^3/(1-\eta-\eta_s)$ | $\kappa\sigma$ | $\kappa\sigma/\sqrt{1-\eta-\eta_s}$ | $a/\sigma$ | $b/\sigma$ | $\Gamma$ | $\xi$ |
|---|---|---|---|---|---|---|---|---|---|---|---|
| A1 | 100 | 10 | 0.1 | 65000 | 41.7 | 1.50 | 1.93 | 2.15 | 1.44 | 1.32 | 0.06 |
| A2 | 100 | 20 | 0.1 | 80000 | 51.3 | 1.66 | 2.15 | 2.15 | 1.44 | 0.93 | 0.18 |
| A3 | 200 | 10 | 0.1 | 115000 | 73.8 | 1.99 | 2.57 | 2.15 | 1.44 | 1.84 | 0.03 |
| A4 | 200 | 20 | 0.1 | 130000 | 83.4 | 2.12 | 2.73 | 2.15 | 1.44 | 1.39 | 0.09 |
| B1 | 100 | 10 | 0.2 | 65000 | 244 | 2.11 | 4.68 | 1.71 | 1.10 | 1.12 | 0.06 |
| B2 | 100 | 20 | 0.2 | 80000 | 300 | 2.34 | 5.19 | 1.71 | 1.10 | 0.76 | 0.19 |
| B3 | 200 | 10 | 0.2 | 115000 | 431 | 2.80 | 6.22 | 1.71 | 1.10 | 1.38 | 0.03 |
| B4 | 200 | 20 | 0.2 | 130000 | 488 | 2.98 | 6.61 | 1.71 | 1.10 | 1.01 | 0.10 |
| C1 | 100 | – | 0.1 | 50000 | 32.1 | 1.32 | – | 2.15 | – | 1.94 | – |
| C2 | 100 | – | 0.2 | 50000 | 188 | 1.84 | – | 1.71 | – | 1.79 | – |
| C3 | 200 | – | 0.1 | 100000 | 64.2 | 1.87 | – | 2.15 | – | 2.38 | – |
| C4 | 200 | – | 0.2 | 100000 | 376 | 2.60 | – | 1.71 | – | 1.96 | – |
| D1 | – | 10 | – | 40000 | 195 | – | 4.18 | – | 7.10 | – | 0.01 |
| D2 | – | 10 | – | 40000 | 19 | – | 1.31 | – | 15.34 | – | 0.001 |

## 3 Results from Primitive Model Simulations

We have simulated globally electroneutral binary colloidal mixtures in a cubic box of edge length $L$ with periodic boundary conditions in all three Cartesian directions. The MD simulation method used here is the same as in Refs.[65,66,73,74]. In order to handle the long-ranged Coulomb interactions[75], we use the Lekner summation method,[76–78] which takes the real-space particle coordinates as its only input. All simulations were carried out at room temperature $T$=293 K, solvent permittivity $\varepsilon$=80, and the big macroion diameter $\sigma$=100 nm. For all runs the small macroion packing fraction $\eta_s$ was more than ten times smaller than that of the big macroions, $\eta$, and the Bjerrum length was equal to $\lambda_B$=0.0071 $\sigma$.

We produced four different series of simulation runs: the low $\eta$ binary colloid runs $A_i$, the high $\eta$ binary colloid runs $B_i$,

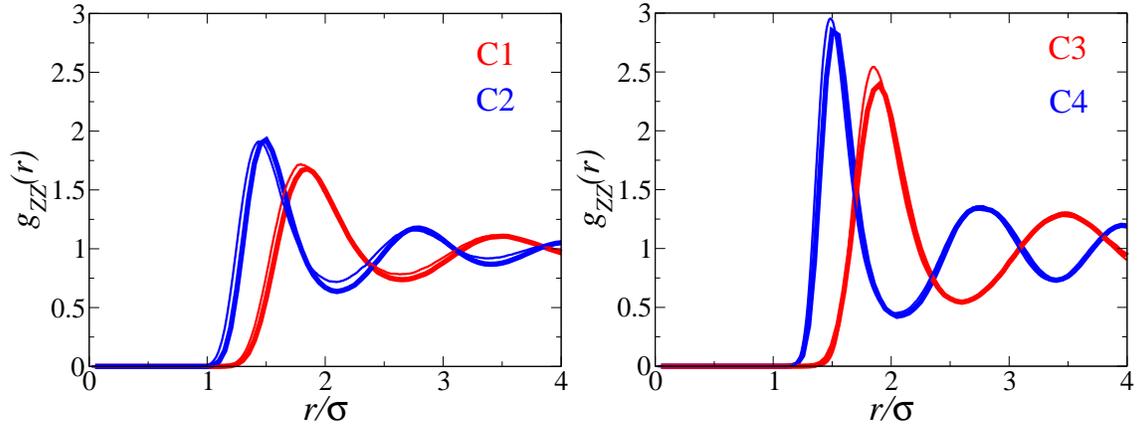

Fig. 3 (Color in online) Pair correlations $g_{ZZ}(r)$ for the runs C1-C4. Thin lines are for the CGM, thick lines are for the PM simulations.

big macroion runs $C_i$ in the absence of small macroions, and small macroion runs $D_j$ in the absence of big macroions, where $i=1,\ldots,4$, and $j=1,2$. Simulation parameters for these runs are collected in Table 1. For the A- and B-series, $N_Z = 500$ and $N_z = 1500$; for the C-series, $N_Z = 500$ and $N_z = 0$; and for the D-series, $N_Z = 0$ and $N_z = 4000$.

A representative snapshot from the simulation box is pictured in Figure 1 for run B2. The color gradient from blue to green along the z-axis indicates the macroion altitude in the simulation box. The macroions are depicted as spheres and the counterions as small red dots.

The main quantities of interest in all runs listed in Table 1 are the pair correlation functions $g_{ij}(r)$, namely $g_{ZZ}(r)$ and $g_{zz}(r)$ for big and small macroions, respectively, and $g_{Zz}(r)$ for cross correlations between big and small macroions. The calculated results for $g_{ij}(r)$ are collected in Figures 2–4, and compared to the prediction of the coarse-grained model (CGM) developed by one of us (AD)[69,71]. This model is based on a sequential coarse-graining scheme, assuming linear-response and mean-field approximations for the small macroion response to the interaction between big and small macroions. The resulting effective pair potentials in the CGM have the Yukawa form with a modified screening constant.

As seen from Figure 2 (first row), there is a systematic deviation between the PM and CGM results for the low $\eta$ runs A1-A4: the height of the first maximum in $g_{ZZ}(r)$ and $g_{Zz}(r)$ in the CGM is higher than in the PM. The height of the first maximum in $g_{zz}(r)$ is practically the same in the PM and CGM data for the runs A1 and A2, whereas, the CGM data underestimate the height of the maximum in the PM for the runs A3 and A4, where the big macroion charge $Z$ is twice as large as in the runs A1 and A2. This observation clearly indicates the strong influence of the big macroion charge on the small-small correlations.

A similar tendency is visible for the high $\eta$ runs B1-B4 in Figure 2 (second row). Here again, the big-big and big-small macroion pair correlations in the CGM show larger first maximum heights, and the small-small pair correlations become less accurate for the highly charged big macroions. Note that the discrepancy between CGM and PM simulation data for $g_{zz}(r)$ in Figure 2 becomes smaller for the lower values of big-big macroion coupling parameter $\Gamma$. For the A-series and B-series simulations the runs A2 and B2 have the lowest $\Gamma$, correspondingly.

Another interesting finding is that in the PM simulations the peak amplitude of the cross-interaction big-small correlation function $g_{Zz}(r)$ is smaller than that of $g_{ZZ}(r)$ and $g_{zz}(r)$. This feature is uncommon for almost symmetric repulsive mixtures, which usually exhibit increasing peak amplitude from $g_{zz}(r)$ to $g_{Zz}(r)$ to $g_{ZZ}(r)$. In contrast to the PM data, the CGM results follow such monotonic behavior.

To understand the origin of the observed discrepancies between theoretical predictions and simulation results, we examine the non-binary systems C1-C4 and D1–D2 for the big and small macroions, respectively. Figure 3 for the runs C1–C4 reveals that the PM and CGM predicted pair correlations $g_{ZZ}(r)$ are practically the same, except for a small difference in the height of the first maximum for run C3. Figure 4 also proves that in colloidal systems with no big macroions, theory and simulation data for $g_{zz}(r)$ are practically identical. This agreement shows that the CGM is accurate for one-component systems of colloids, regardless of macroion size and charge. It follows, therefore, that in binary mixtures the big colloids perturb the distribution of the small colloids and counterions in a way that the CGM theory does not fully capture. It should be kept in mind that the CGM assumes spherically symmetric counterion and small macroion distributions around the big macroions even when two big macroions closely approach each other.

To access the anisotropy of the counterion cloud around the macroions, we define the averaged counterion density field $\rho_c(\vec{r}) \equiv \rho_c(\vec{r}, \{\vec{r}_i^{(Z)}, \vec{r}_j^{(z)}\})$ which parametrically depends on the fixed macroion positions $\{\vec{r}_i^{(Z)}, \vec{r}_j^{(z)}, i = 1,...,N_Z; j = 1,...N_z\}$. As usual, we define

$$\rho_c(\vec{r}) = \left\langle \sum_{\ell=1}^{N_c} \delta\left(\vec{r} - \vec{r}_\ell^{(c)}\right) \right\rangle_c, \qquad (2)$$

performing a canonical counterion average $\langle \cdots \rangle_c$ for fixed macroion positions. In Figure 5 we show the counterion density field around the big macroions for run A2 as obtained by an additional average over typical macroion positions. This field is directionally resolved, the direction being fixed by neighboring macroions. The resulting $\rho_c(\theta)$ shows that the counterions are mainly located between two big macroions where the direction angle $\theta = 0$. Next one can similarly average the density field $\rho_z(\vec{r})$ of the small macroions around fixed big macroions. The directionality of this field $\rho_z(\theta)$ is also shown in Figure 5 and shows a peak at a finite direction angle $\theta \approx 5\pi/12$.

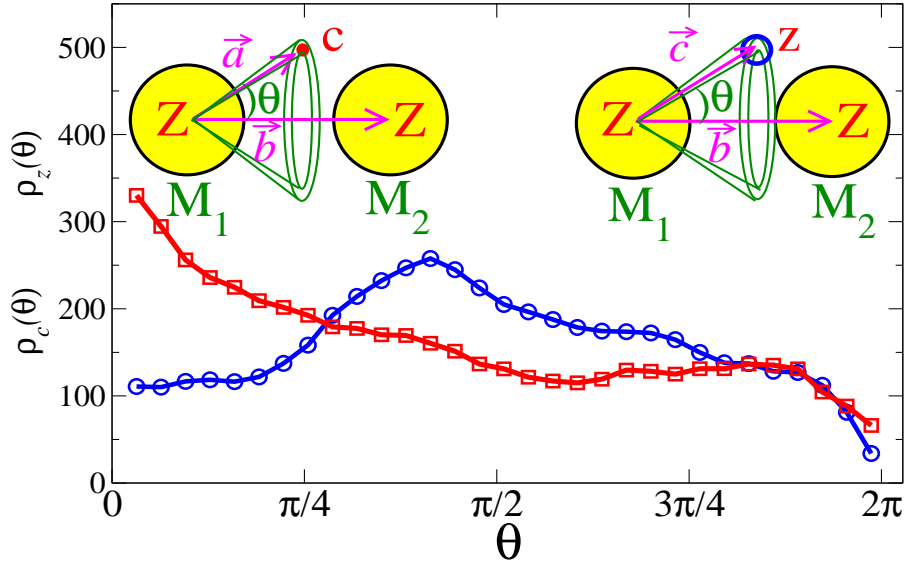

Fig. 5 (Color in online) Simulation results for the non-normalized density field of counterions $\rho_c(\theta)$ (red line) and the small macroions $\rho_z(\theta)$ (blue line) around the big macroions for run A2 from Table 1. The insets explain the directionality of this field for a pair of big macroions $M_1$ and $M_2$. Each distribution corresponds to the number of corresponding particles in the 3D conical shell of width 0.1 radians around the big macroion $M_1$ with a proper angular normalization factor $1/\sin\theta$. A sample counterion is shown as a red dot in the left inset and a sample small macroion as a hollow blue circle in the right inset. The vectors $\vec{a}$ and $\vec{b}$ are defined as $\vec{a} = \vec{r}_\ell^{(c)} - \vec{r}_{M_1}^{(Z)}$, $\vec{b} = \vec{r}_{M_2}^{(Z)} - \vec{r}_{M_1}^{(Z)}$, and therefore the direction angle is $\theta = \mathrm{acos}\left(\vec{a}\cdot\vec{b}/|\vec{a}||\vec{b}|\right)$. The vector $\vec{c}$ is defined as $\vec{c} = \vec{r}_i^{(z)} - \vec{r}_{M_1}^{(Z)}$, and therefore the direction angle in this case is $\theta = \mathrm{acos}\left(\vec{c}\cdot\vec{b}/|\vec{c}||\vec{b}|\right)$. In the calculation of $\rho_\alpha(\theta)$ ($\alpha = z,c$) we accounted for only the big macroion pairs with the separation distance $b < 1.8\sigma$, counterions with $a < b/2$, and small macroions with $c < b/2$.

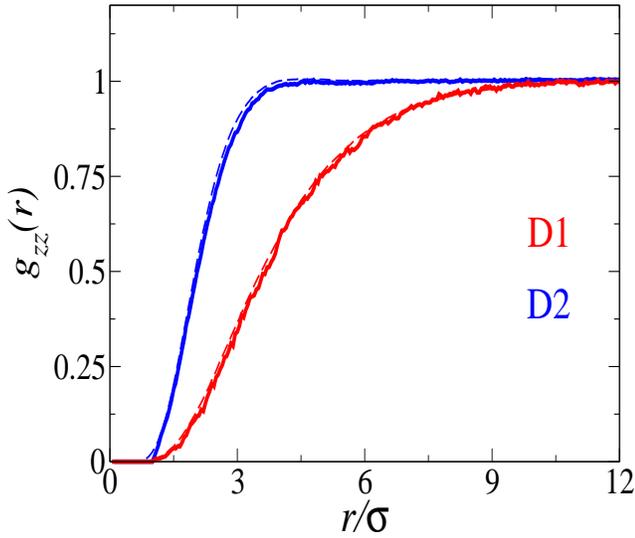

Fig. 4 (Color in online) Pair correlations $g_{zz}(r)$ for the runs D1 and D2. Thin lines are for the CGM, thick lines are for the PM simulations.

## 4 The role of entropic forces in binary mixtures

The excluded volume of the macroions generates entropic forces arising from the contact density of counterions at the macroion surface. The entropic force acting on the $i$-th macroion of species $\alpha$ at the position $\vec{r}_i^{(\alpha)}$ with $i \in 1,...,N_\alpha$ ($\alpha = Z,z$) is defined as [73,74,79–82].

$$\vec{F}_{ent}^{(\alpha)}(\vec{r}_i^{(\alpha)}) = -k_B T \int_{S_i^{(\alpha)}} d\vec{f} \, \rho_c(\vec{r}), \qquad (3)$$

where $\vec{f}$ is a surface normal vector pointing outwards from the macroion's core and $S_i^{(\alpha)}$ is the surface of the hard core of the $i$-th macroion centered around $\vec{r}_i^{(\alpha)}$ with diameter $(\sigma_\alpha + \sigma_\beta)/2$. The entropic force, usually neglected in weakly charged macroion systems, strongly modifies the macroion interactions in highly charged and dense colloidal systems.

Likewise, the canonically averaged electrostatic force acting on the $i$-th macroion of species $\alpha$ is defined as

$$\vec{F}_{elec}^{(\alpha)}(\vec{r}_i^{(\alpha)}) = \left\langle \sum_{\beta = Z,z,c} \sum_{j=1}^{N_\beta} \vec{F}^{(\alpha\beta)}(\vec{r}_i^{(\alpha)} - \vec{r}_j^{(\beta)})\left(1 - \delta_{\alpha\beta}\,\delta_{ij}\right)\right\rangle_c, \qquad (4)$$

where $\alpha = Z,z$. The Kronecker delta functions in this expression nullify the self-interaction of macroions. Clearly, in Eq.(4) the electrostatic pair interaction forces $\vec{F}^{(\alpha\beta)}$ are defined as

$$\vec{F}^{(\alpha\beta)}(\vec{r}_{ij}) = -\vec{\nabla}_{\vec{r}_{ij}} V^{(\alpha\beta)}(r_{ij}) = \frac{1}{4\pi\varepsilon_0}\frac{q^{(\alpha)}q^{(\beta)}}{\varepsilon r_{ij}^2}\frac{\vec{r}_{ij}}{r_{ij}}, \qquad (5)$$

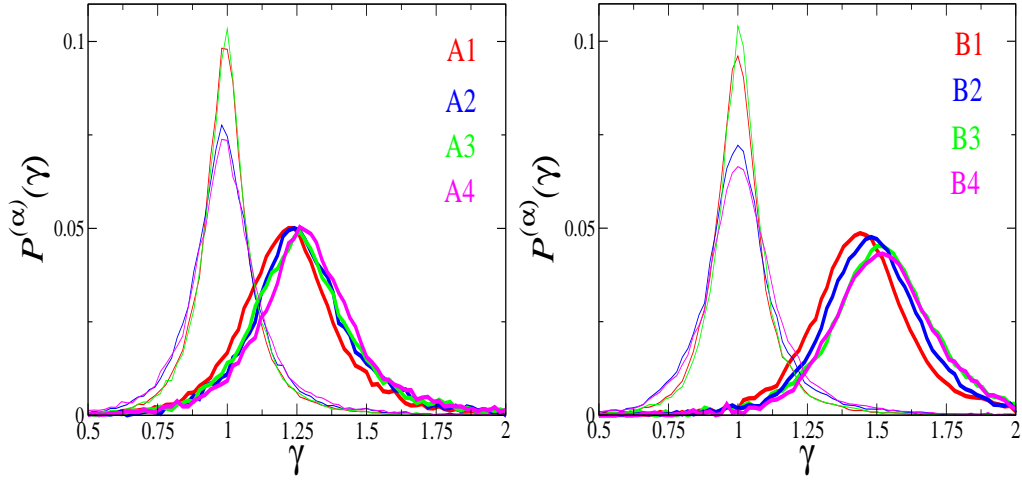

**Fig. 6** (Color in online) Normalized and averaged distribution $P^{(\alpha)}(\gamma)$ of the force-ratio factor $\gamma$ for big ($\alpha = Z$ and thick lines) and small ($\alpha = z$ and thin lines) macroions for the runs A1-A4 and B1-B4.

for $r_{ij} > \sigma_{\alpha\beta}$, where $\vec{r}_{ij} = \vec{r}_i^{(\alpha)} - \vec{r}_i^{(\beta)}$.

The contribution of the entropic forces acting on the macroions of species $\alpha = Z, z$ can be evaluated by the force-ratio factor

$$f^{(\alpha)} = \frac{1}{N_\alpha} \sum_{i=1}^{N_\alpha} f_i^{(\alpha)}, \quad f_i^{(\alpha)} = \frac{|\vec{F}_{elec}^{(\alpha)}(\vec{r}_i^{(\alpha)}) + \vec{F}_{ent}^{(\alpha)}(\vec{r}_i^{(\alpha)})|}{|\vec{F}_{elec}(\vec{r}_i^{(\alpha)})|}. \quad (6)$$

Obviously, $f_i^{(\alpha)} > 1$ implies that the entropic force acting on the $i$-th macroion of species $\alpha$ is aligned with the electrostatic force, while $f_i^{(\alpha)} < 1$ means anti-alignment. The averaged force-ratio distribution,

$$P^{(\alpha)}(\gamma) = \left\langle \left\langle \delta\left(\gamma - f^{(\alpha)}\right) \right\rangle_c \right\rangle_m, \quad (7)$$

where $\langle \cdots \rangle_m$ is a full canonical average over both macroion species, is shown in Figure 6 for the runs A1–A4 and B1–B4. For the small macroions $P^{(z)}(\gamma)$ is maximal at $\gamma=1$ regardless of the packing fraction $\eta$. This condition means that, in most cases, the contact counterion density at the small macroion surface is spherically symmetric and experiences no distortion from the electric field of the neighboring macroions. It is also apparent that the force-ratio distribution is affected by the small macroion charge $z$: the higher the charge $z$, the broader the distribution $P^{(z)}(\gamma)$, and its height at the maximum becomes smaller.

The big macroion force-ratio distribution $P^{(Z)}(\gamma)$ is also shown in Figure 6. The distribution is centered around $\gamma=1.25$ for the low $\eta$ runs A1-A4, and around $\gamma=1.5$ for the high $\eta$ runs B1-B4. These observations imply that the distortion of the counterion contact density at the big macroion surface (and associated with it the bolstering effect of the entropic forces) increases with the macroion packing fraction $\eta$. Such bolstering effect, ignored in the CGM model, might explain the discrepancies observed between the PM and CGM data in Figure 2.

Figure 6 also reveals that the position of the maximum in $P^{(Z)}(\gamma)$ systematically shifts to higher values of $\gamma$ from run A1 to run A4, and from run B1 to run B4. These shifts can be explained by the increase of the available-volume-corrected counterion density $n_c/(1-\eta-\eta_s)$ along these runs as seen from Table 1, which intensifies the contribution of the entropic forces.

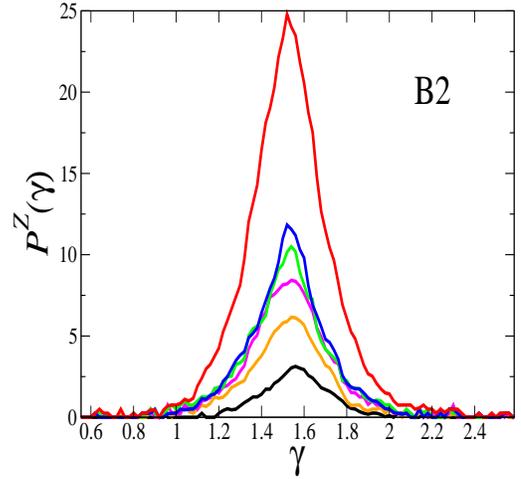

**Fig. 7** (Color in online) Non-normalized and averaged distribution $P^{(Z)}(\gamma)$ of the force-ratio factor $\gamma$ for the big macroions in run B2 for different macroion-macroion separation distances $r$. From bottom to top: black line- $1.0 \leq r/\sigma < 1.1$, orange line- $1.1 \leq r/\sigma < 1.2$, pink line- $1.2 \leq r/\sigma < 1.3$, green line- $1.3 \leq r/\sigma < 1.4$, blue line- $1.4 \leq r/\sigma < 1.5$, red is for $r/\sigma \geq 1.5$.

In Figure 7 the distribution $P^{(Z)}(\gamma)$ is resolved according to the macroion distance. The force-ratio shift for the highly charged macroion has practically no dependence on the macroion-macroion separation distance $r$. This insensitivity is a manifestation of the fact that the counterion cloud distortion is a robust effect and is strong even at the separation distances comparable with the average macroion separation distance $a$ in the system.

## 5 Optimal effective pair interaction between macroions

## 5.1 Fitting of the PM macroion forces with non-additive Yukawa forces

In the previous section it was found that the CGM did not completely reproduce the PM simulation results for the pair correlation functions. An improvement can be sought in the implementation of non-additive Yukawa pair interaction potentials between the species $\alpha$ and $\beta$ with $\alpha, \beta \in \{Z_Y, z_Y\}$[24,65,68],

$$V_Y^{(\alpha\beta)}(r_{ij}) = \begin{cases} \infty, & \text{for } r_{ij} \leq \sigma_{\alpha\beta}, \\ q^{(\alpha)}q^{(\beta)} e^{-\kappa_Y r_{ij}} \frac{1+(1-\delta_{\alpha\beta})\Delta_Y}{4\pi\varepsilon_0\varepsilon r_{ij}}, & \text{for } r_{ij} > \sigma_{\alpha\beta}, \end{cases} \quad (8)$$

where $q^{(\alpha)}$ and $q^{(\beta)}$ with $\alpha, \beta = Z_Y, z_Y$ denote effective Yukawa charges for the macroions, $\vec{r}_{ij} = \vec{r}_j^{(\alpha)} - \vec{r}_i^{(\beta)}$ with $i \in 1,...,N_\alpha$ ($\alpha = Z_Y, z_Y$) and $j \in 1,...,N_\beta$ ($\beta = Z_Y, z_Y$) is the distance between the two effective Yukawa charges, $\Delta_Y$ is the non-additivity parameter between the big and small macroions, $\kappa_Y$ is an effective Yukawa inverse screening length, and $\delta_{\alpha\beta}$ is the Kronecker delta. The corresponding interaction forces between the species $\alpha$ and $\beta$ are

$$\vec{F}_Y^{(\alpha\beta)}(\vec{r}_{ij}) = \frac{q^{(\alpha)}q^{(\beta)}}{\varepsilon r_{ij}^2} e^{-\kappa_Y r_{ij}} \left(\frac{1}{r_{ij}} + \kappa_Y\right)\left(1 + (1-\delta_{\alpha\beta})\Delta_Y\right)\vec{r}_{ij} \quad (9)$$

The optimal parameters for the effective Yukawa potential are deduced from the best fit between the PM and Yukawa forces acting on the macroions during the runs. For this purpose, for the runs in Table 2 we stored entropic $\vec{F}_{ent}^{(\alpha)}(\vec{r}_i^{(\alpha)})$ and electrostatic $\vec{F}_{elec}^{(\alpha)}(\vec{r}_i^{(\alpha)})$ forces acting on the macroions during the PM simulations [see Eqs.(3) and (4)]. In total, we collected $N_{conf} = 100$ independent configurations for each run, and each configuration contained the set of $\{\vec{F}_{elec}^{(\alpha)}(\vec{r}_i^{(\alpha)})\}$, $\{\vec{F}_{ent}^{(\alpha)}(\vec{r}_i^{(\alpha)})\}$, and $\{\vec{r}_i^{(\alpha)}\}$ data, where $i \in 1,...,N_\alpha$ ($\alpha = Z_Y, z_Y$).

Then, using the stored macroion positions $\{\vec{r}_i^{(\alpha)}\}$, we calculated the devised Yukawa forces $\vec{F}_Y^{(\alpha)}(\vec{r}_i^{(\alpha)})$ acting on the macroions,

$$\vec{F}_Y^{(\alpha)}(\vec{r}_i^{(\alpha)}) = \sum_{\beta=Z_Y,z_Y} \sum_{j=1}^{N_\beta} \vec{F}_Y^{(\alpha\beta)}(\vec{r}_{ij})\left(1 - \delta_{\alpha\beta}\delta_{ij}\right), \quad (10)$$

and, for each run, performed the following least-square fitting procedure,

$$\min\left[\left\langle \sum_{\alpha=Z,z} \sum_{i=1}^{N_\alpha} \left(\vec{F}_Y^{(\alpha)}(\vec{r}_i^{(\alpha)}) - \vec{F}_{elec}^{(\alpha)}(\vec{r}_i^{(\alpha)}) - \vec{F}_{ent}^{(\alpha)}(\vec{r}_i^{(\alpha)})\right)^2 \right\rangle_m\right] \quad (11)$$

to get the optimal fit values for the effective big macroion charge $Z_Y^{(opt)}$, the effective small macroion charge $z_Y^{(opt)}$, the non-additivity parameter $\Delta_Y^{(opt)}$, and the inverse screening length $\kappa_Y^{(opt)}$.

The obtained Yukawa fitting coefficients for the runs A1-A4 and B1-B4 are summarized in Table 2. As a result, the non-additivity parameter $\Delta_Y^{(opt)}$ is small and varies between -0.003 and 0.079. Effective big macroion charges are always larger than their bare charges, whereas effective small macroion charges are either smaller or almost equal to their bare charges.

Optimal fitting parameters, presented in Table 2 were used in the binary Yukawa mixture simulations to calculate macroion-macroion pair correlations $g_{ij}^Y(r)$, which are presented in Figures 8 and 9. Here the PM simulation data are given in red and the Yukawa mixture data are given in green. For the low $\eta$ runs A1–A4 in Figure 8, the Yukawa mixture data are close to the PM simulation data for the cross macroion-macroion pair correlation line $g_{Zz}(r)$, but they overestimate the PM simulation data for $g_{ZZ}(r)$ and underestimate the PM simulation data for $g_{zz}(r)$. A similar tendency is seen for the high $\eta$ runs B1–B4 in Figures 9. Again, the effective Yukawa mixture data are in good agreement with the PM data for $g_{Zz}(r)$, but they overestimate the PM data for $g_{ZZ}(r)$ and underestimate the PM data for $g_{zz}(r)$. In total, the sequence of peak amplitudes in $g_{zz}$, $g_{zZ}$, and $g_{ZZ}$ is always monotonic for the effective Yukawa model, while it is non-monotonic in the full PM. This might also explain why an effective Yukawa picture in the CGM is not sufficient to get the non-monotonicity.

Table 2 Optimal fit values for $Z_Y^{(opt)}$, $z_Y^{(opt)}$, $\Delta_Y^{(opt)}$, and $\kappa_Y^{(opt)}$ for the binary and non-additive Yukawa system.

| Run | $Z_Y^{(opt)}$ | $z_Y^{(opt)}$ | $\Delta_Y^{(opt)}$ | $\kappa_Y^{(opt)}\sigma$ |
|---|---|---|---|---|
| A1 | 128.24 | 10.04 | -0.003 | 1.44 |
| A2 | 129.36 | 19.26 | 0.009 | 1.52 |
| A3 | 247.85 | 10.19 | -0.008 | 1.81 |
| A4 | 246.22 | 19.06 | -0.001 | 1.85 |
| B1 | 124.79 | 8.64 | 0.079 | 1.55 |
| B2 | 139.88 | 17.30 | 0.016 | 1.64 |
| B3 | 237.99 | 8.74 | 0.076 | 1.83 |
| B4 | 238.70 | 17.06 | 0.024 | 2.15 |

## 5.2 Matching PM macroion forces with non-additive Yukawa and attractive Gaussian forces

We tried to find alternative Yukawa-like models for the best fitting of the PM simulation results for the pair correlations $g_{zz}(r)$ in Figures 8 and 9. As a first attempt, we designed a modified Yukawa model with separate inverse screening lengths for the big-big, small-small, and big-small macroion interactions, respectively. This model, however, failed to improve the fitting of the PM simulation data. As a second attempt, we used a double-repulsive Yukawa model with two interaction force terms for the effective macroion-macroion interactions. This model also did not significantly improve the fitting of the PM simulation results.

A model that did prove successful incorporates an attractive and short-ranged Gaussian potential between the small macroions,

$$\frac{U_G(r)}{k_B T} = -A_G \exp\left(-\frac{(r-b_G)^2}{s_G^2}\right) \quad (12)$$

on top of effective Yukawa repulsion as assumed in section 5.1. The corresponding Gaussian attractive force,

$$\frac{\vec{F}_G(r)\sigma}{k_B T} = -2A_G \frac{r-b_G}{s_G^2} \exp\left(-\frac{(r-b_G)^2}{s_G^2}\right)\frac{\vec{r}}{r}, \quad (13)$$

between the small macroions is capable to raise the height of the first maximum in $g_{zz}(r)$. Optimal fit parameters for this Yukawa-Gaussian model are collected in Table 3 for the runs A1–A4 and B1–B4. From Table 2 and 3 it is evident that the fit charge $\tilde{Z} < Z_Y^{(opt)}$ and the fit inverse screening length $\tilde{\kappa} < \kappa_Y^{(opt)}$.

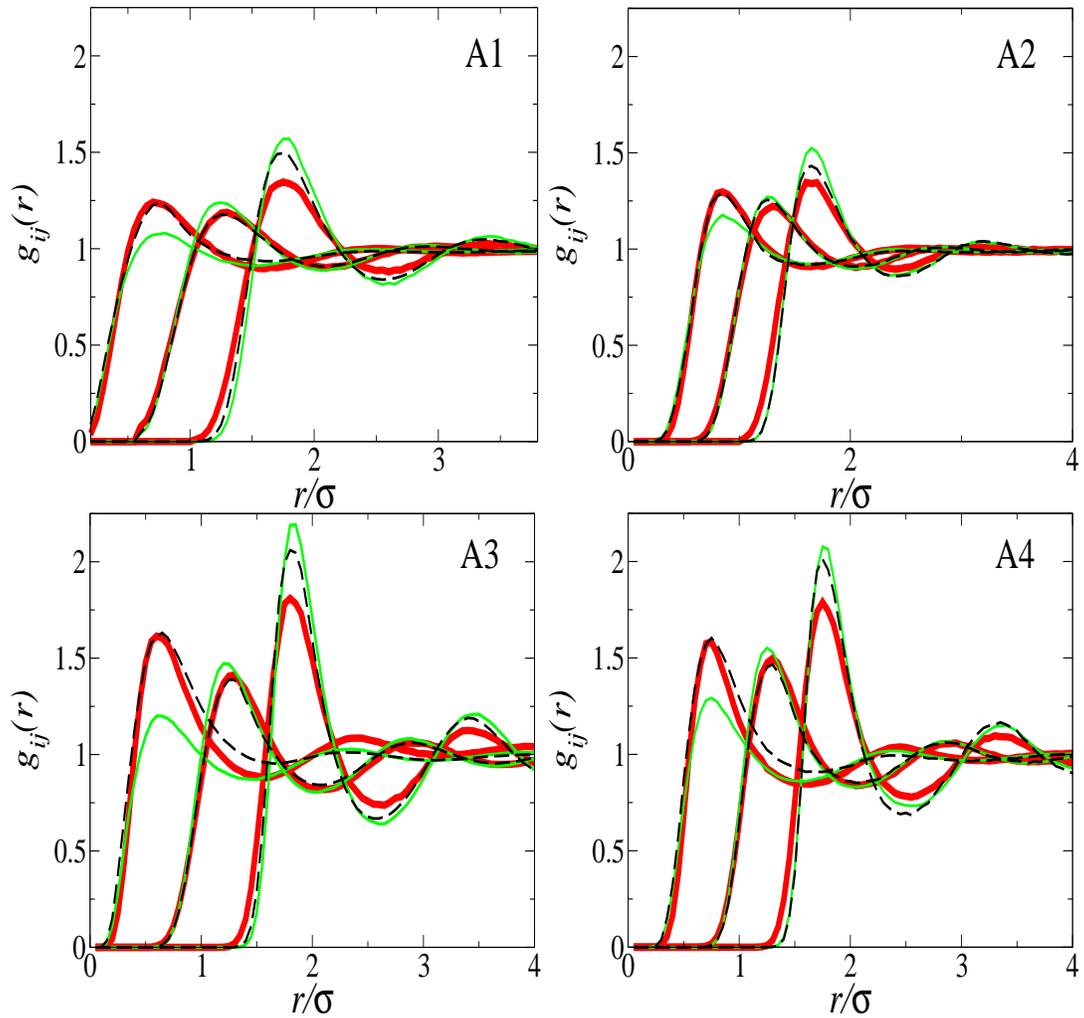

Fig. 8 (Color in online) Pair correlations $g_{ij}(r)$ for the low $\eta$ runs A1-A4. Red lines are for the PM results, green lines are results for the binary Yukawa model with parameters from Table 2, and dashed black lines are for the binary Yukawa with Gaussian attraction model with parameters from Table 3.

The Gaussian term (Eq.(12) for the runs A1–A4 and B1–B4 is shown in Figure 10a. The depth of the attractive well varies between -0.2$k_B T$ and -0.6$k_B T$. The total interaction potentials between the small macroions, plotted in Figure 10b, show that the Gaussian attraction is weaker than the Yukawa repulsion, such that the total effective interaction stays mainly repulsive. For the runs A1, A2, B1, and B3 with a low charge $z$, the total interaction potential has a plateau at the distance $r \approx \sigma/2$.

Binary Yukawa-Gaussian simulation results for $g_{ij}(r)$ with the optimal fit parameters from Table 3 are shown as black dashed lines in Figures 8 and 9. Indeed, the incorporation of the attractive Gaussian potential into the small-small macroion interaction improves the fitting of the PM simulation data for $g_{zz}(r)$ significantly. Moreover, the non-monotonicity in the peak amplitude for $g_{zz}$, $g_{zZ}$, and $g_{ZZ}$ is reproduced in the Yukawa-Gaussian model.

Finally, let us propose a simple picture for the physical origin of the effective $z-z$ attraction. It is intuitive to consider the counterion density cloud, which is essentially dictated and governed by the big macroions. This screening cloud can be obtained from Figure 5 and is sketched in Figure 11 for a triangular triplet of big macroions. The small macroions occupy the voids between the macroions, as revealed by $\rho(\theta)$ in Figure 5 and sketched in Figure 11. Now small macroions opposed to the screening cloud of counterions are effectively mutually attracted via the counterion cloud. This happens in particular at a typical distance between two small macroions at around $\sigma/2$ where the shifted Gaussian attraction is minimal.

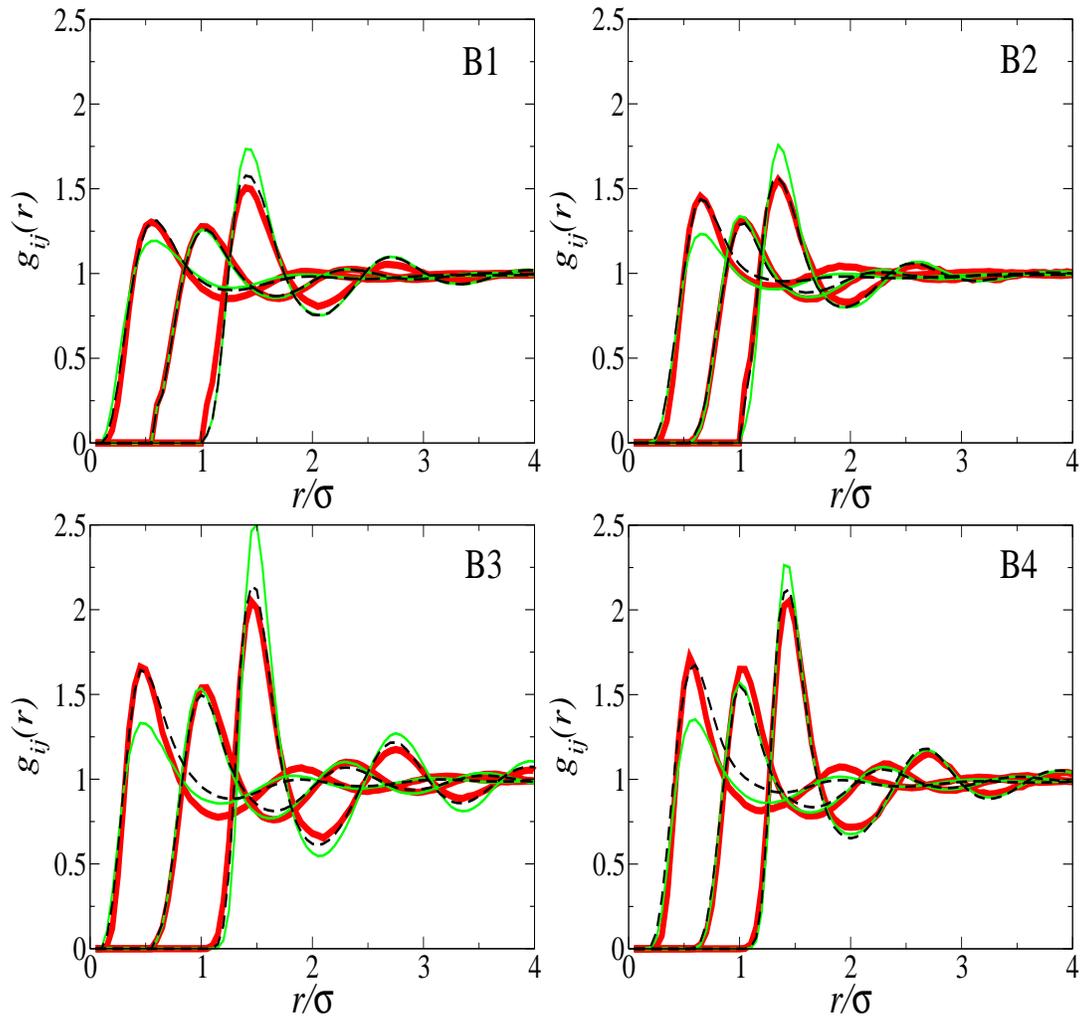

Fig. 9 (Color in online) Pair correlations $g_{ij}(r)$ for the low $\eta$ runs B1-B4. Red lines are for the PM results, green lines are results for the binary Yukawa model with parameters from Table 2, and dashed black lines are for the binary Yukawa with Gaussian attraction model with parameters from Table 3.

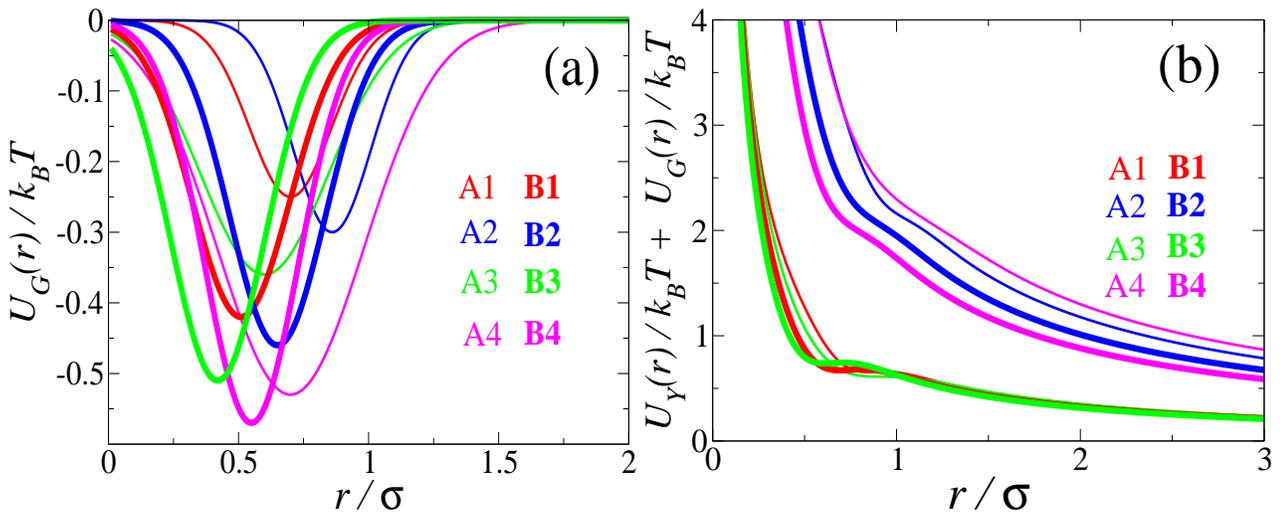

Fig. 10 (Color in online) (a) The Gaussian attraction between the small macroions added to the binary Yukawa model. (b) Total interaction potential within the Yukawa-Gaussian model between the small macroions defined as the sum of the repulsive Yukawa and attractive Gaussian potentials. Thin lines with four different colors are for the runs A1-A4, thick lines with four different colors are for the runs B1-B4.

Table 3 Parameters of the different runs of the binary Yukawa-Gaussian system simulation. The Coulomb coupling parameter between the big macroions is $\Gamma_{YG} = \bar{Z}^2 \exp(-\bar{k}a)\lambda_B/a$, and between the small macroions is $\xi_{YG} = \bar{z}^2 \exp(-\bar{k}b)\lambda_B/b$, where the subscript YG refers to the Yukawa-Gaussian model. The meanings of the other quantities in the first row are explained in the text.

| Run | $\bar{Z}$ | $\bar{z}$ | $\bar{\Delta}$ | $\bar{k}\sigma$ | $A_G$ | $b_G$ | $s_G$ | $\Gamma_{YG}$ | $\xi_{YG}$ |
|---|---|---|---|---|---|---|---|---|---|
| A1 | 121.1 | 9.85 | 0.027 | 1.57 | 0.25 | 0.70 | 0.05 | 1.66 | 0.05 |
| A2 | 121.35 | 18.17 | 0.018 | 1.64 | 0.30 | 0.86 | 0.05 | 1.44 | 0.15 |
| A3 | 233.43 | 9.93 | 0.017 | 1.88 | 0.36 | 0.60 | 0.12 | 3.17 | 0.03 |
| A4 | 245.93 | 19.09 | -0.050 | 1.82 | 0.53 | 0.70 | 0.16 | 4.03 | 0.13 |
| B1 | 123.70 | 9.52 | -0.043 | 1.71 | 0.42 | 0.51 | 0.071 | 3.45 | 0.09 |
| B2 | 134.62 | 16.84 | -0.023 | 2.10 | 0.46 | 0.65 | 0.067 | 2.07 | 0.18 |
| B3 | 232.66 | 9.40 | -0.020 | 2.29 | 0.51 | 0.42 | 0.066 | 4.49 | 0.05 |
| B4 | 235.78 | 15.72 | -0.040 | 2.35 | 0.57 | 0.55 | 0.065 | 4.15 | 0.12 |

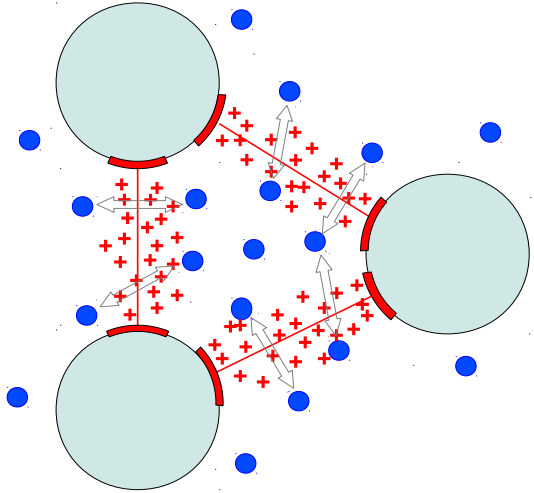

Fig. 11 (Color in online) Schematic picture explaining the origin of the effective attraction between the small macroions. Transparent double-ended arrows point to the small macroions, which are effectively attracted to each other via the screening counterion cloud between the macroions.

## 6 Conclusions

To summarize, we have calculated the pair correlations in strongly asymmetric like charge macroion mixtures using the primitive model with explicit counterions. We have compared our data to a coarse-grained theory proposed earlier and found agreement only for small Coulomb couplings. While we found in the simulations that the big-small correlation peak was smaller than that of the big-big and small-small correlations, this could not be reproduced by the theory. In our simulations we are dealing with effective many-body forces between the macroions. If these are fitted to effective pairwise interactions, optimal pair potentials can be extracted. We demonstrated here that with these optimal pairwise interactions the full simulation data for the pair correlations can be reproduced. This indicates that effective triplet interactions are small[83,84]. However, the optimal effective pairwise interactions are not just a non-additive repulsive Yukawa interactions but strikingly also involve an additional shifted attractive Gaussian potential in the effective small-small interactions. We add a remark here that our optimal pair interactions are different to those which exactly reproduce the pair correlations according to the theorem of Henderson[85,86]. Our potentials also embody many-body forces while the latter are substitute potentials to reproduce the pair correlations exactly.

Our simulation results provide benchmark data to test future theories for the macroion pair structure in strongly asymmetric mixtures. These should incorporate also entropic interactions since we found that those contribute significantly to the total interactions.

Moreover if it comes to an actual comparison to experimental data, the size- and charge polydispersity of the sample needs to be checked and possible incorporated in the theoretical description. In future simulations, a finite concentration of added salt with microscopic coions[66] should be included. Furthermore an explicit solvent should be considered, see e.g.[87–91].

Finally, in our model we neglected van der Waals attractions. If these are taken into account, they will compete with the Coulomb forces[92], which might lead to new structural ordering effects. To explore this for binary mixtures is a promising field of future research.

## Author Contributions

All authors contributed equally to this manuscript by developing the theoretical formalism, performing the analytic calculations and performing the numerical simulations. All authors participated in writing the manuscript.

## Conflicts of interest

There are no conflicts to declare.

## Acknowledgments

H.L. thanks the financial support from the DFG within project LO418/23-1. A.D. thanks the financial support from the NSF within the Grant No. DMR-1928073. E.A. thanks the financial support from the Ministry of Science and Higher Education of the Russian Federation (State Assignment No. 075-01056-22-00).